# Large Range Manipulation of Exciton Species in Monolayer WS$_2$

**Ke Wei,**[1,5,6] **Yu Liu,**[1,5,6] **Hang Yang,**[3] **Xiangai Cheng,**[1,2,5,6] **Tian Jiang,**[1,2,4,5,6*]

[1]*College of Opto-Electronic Science and Engineering, National University of Defense Technology, Changsha, 410073, China.*

[2]*State Key Laboratory of High Performance Computing, National University of Defense Technology, Changsha, 410073, China.*

[3]*College of Science, National University of Defense Technology, Changsha, 410073, China.*

[4]*State Key Laboratory of Low-Dimensional Quantum Physics, Department of Physics, Tsinghua University, Beijing, 10000, China.*

[5]*Hunan Provincial Key Laboratory of High Energy Laser Technology, Changsha, 410073, China.*

[6]*Hunan Provincial Collaborative Innovation Center of High Power Fiber Laser, Changsha, 410073, China.*

[*]*Corresponding author: tjiang@nudt.edu.cn*

**Absrtact:** Unconventional emissions from exciton and trion in monolayer WS$_2$ are studied by photoexcitation. Excited by 532nm laser beam, the carrier species in the monolayer WS$_2$ are affected by the excess electrons escaping from photoionization of donor impurity, the concentration of which varies with different locations of the sample. Simply increasing the excitation power at room temperature, the excess electron and thus the intensity ratio of excited trion and exciton can be continuously tuned over a large range from 0.1 to 7.7. Furthermore, this intensity ratio can also be manipulated by varying temperature. However, in this way the resonance energy of the exciton and trion show red-shifts with increasing temperature due to electron-phonon coupling. The binding energy of the trion is determined to be ~23meV and independent to temperature, indicating strong Coulomb interaction of carriers in such 2D materials.

## 1. Introduction

The extraordinary properties [1-4] of atomically thin transition-metal dichalcogenides (TMDs) beyond graphene have brought these materials to the forefront of research. With carriers confined to a plane, dielectric screening of electric field is substantially reduced, resulting in a strong Coulomb interaction between electrons and holes. Exciton binding energies in such monolayer TMDs are expected to be several hundred millielectronvolts (meV) [5, 6]. Additionally, the group of TMDs shows a crossover from an indirect bandgap in multilayer to a direct bandgap at the limit of monolayer, resulting in a drastic increase of luminescence quantum efficiency in monolayer TMDs [7]. Furthermore, due to spin-orbit coupling (SOC) [8] and broken inversion symmetry [9], both of the absorption and emission in monolayer TMDs show valley and spin contrasting optical selection rule, providing an unprecedented platform to explore valleytronics. These exotic properties make them promising materials in electronic, optoelectronic and valleytronics applications such as transistors [10], light-emitting diodes [11], and logic gates [12].

Information on elementary excitations is essential to improve the performance of the device. Owing to the strong Coulomb interactions, exciton domains carrier dynamics in 2D semiconductors. The neutral exciton can further combine with other particles to form many-body excitons if the Coulomb interaction is strong enough. An example is monolayer TMDs, in which an exciton usually binding with an additional electron (X+) or hole (X-) to form charged exciton (or trion) [13-15]. This three-body exciton often shows different properties to the neutral exciton, especially in the valley dynamics [16, 17]. Thus the ability to control the carrier species, especially the ratio between trion and exciton, is very important to the application of monolayer TMDs. Precious studies show that the generation of trion is determined by the concentration of excess electron (or hole), which can be control by gate-doping [13, 15, 18], photoionization of impurity [14], substrate [19], functionalization layer [20] and naturally charge transfer in TMDs heterostructure [21]. Trion manipulation in such ways always requires low-temperature limit, extra dopping or heterostructures, which is complex or inconvenient for real application.

Here we demonstrate that the carrier species can be continuously and substantially control by excitation intensity in naturally n-dope monolayer $WS_2$ at room temperature. Excited by 532nm laser beam, the emission intensity ratio between trion and exciton (trion/exciton) can be continuously tuned from 0.1 to 7.7 when the excitation density increases from 0.24kW/cm$^2$ to 204kW/cm$^2$. This drastic change of intensity ratio can be attributed to high donor impurity density in the monolayer $WS_2$. Moreover, the carrier species can also be tuned by varying temperatures. While in this way the resonance energies of both exciton and trion show red-shift with increasing temperature due to electron−phonon coupling. From fitting the PL spectrum, the trion binding energy in monolayer $WS_2$ is found to be about 23meV, compared to previously reported values [5, 15] and slightly increased (<4meV) with excitation intensity but independent to temperature.

**2. Results and discussion**

*2.1 Characterization of monolayer $WS_2$*

Single and few layer $WS_2$ flakes are obtained by mechanical exfoliation from bulk 2H-$WS_2$, which naturally shows n type without gate-doping. Two monolayer flakes on 300nm $SiO_2$ on Si substrates are used for data verification and comparison and identified by optical microscopy, atomic force microscopy (AFM), and Raman spectroscopy. Typical results (Flake 1) are shown Fig.1. AFM height profile shows a step of 0.8nm of the flake, corresponding to a single-layer $WS_2$ [22]. Under 532nm laser excitation, Raman spectrum of the monolayer $WS_2$ is obtained at room temperature, as shown in Fig. 1(b). Rich peaks are found as previous report [23]. Several peaks are symmetry assigned, including the longitudinal acoustic phonons (LA(M)) and its second order mode (2LA(M)), the out-of-plane phonon mode ($A_{1g}(\Gamma)$) and the Raman peak from Si substrate. Here the in-plane phonon mode cannot be distinguished since it is overlapped by the strong 2LA(M) mode.

*2.2 Impurity dependent photoluminescence*

For estimation of the uniformity and the impurity distribution on the monolayer $WS_2$, Spatial photoluminescence mapping (Flake 2) is performed, as shown in Fig. 2. The excited beam (532nm) is focused by the objective (×100) laser beam with size (FWHM) of 0.95μm. Galvanometer scanning is used for PL spatial mapping with spatial resolutions of ~0.1μm. The samples are measured at right angle geometry and the PL is analyzed by spectrometer equipped with a CCD camera. The mapping is performed at room temperature.

Fig. 2(b) shows the total PL intensity map (integrated form 600nm to 650nm) of the area marked in Fig. 2(a). The intensity is nearly the same in the whole flake, indicating a uniform and unstressed single-layer $WS_2$ [24]. Some specific patterns at the center of the flake may be attributed to the residue of tapes. Noticed that at most locations on the flake, the PL spectrum contains two peaks (Fig. 2(d)), which can be well separated by double-Gaussian-peak fitting with resonance energy located around 1.98eV and 2.01eV, respectively. Previous reports show that the high energy peak can be attributed to the radiative recombination from the bandage exciton of monolayer $WS_2$ [5, 15]. While the lower energy peak, as discussed below, has all characteristics of a trion. Thus this low-energy line is attributed to the negative trion since the monolayer $WS_2$ is n type, inheriting from the bulk crystal.

The intensity ratio of the two carriers is different at different locations in the monolayer $WS_2$, as shown in Fig. 2(c and d). Especially at the outside ring of the flake (point 1 and 5), the intensity ratio of exciton/trion varies from 0.14 to 1. Since the trions are only formed in the presence of excess charge, the exciton/trion ratio spatial image reflects the density distribution of the excess charge. Trion is also observed at room temperature using excess gate-doping [15]. Here for ungated monolayers, the existing of trion can be explained by the naturally n-dopped of the monolayer. More importantly, laser illumination on monolayer $WS_2$, except to excite exciton, is also expected to photoionize carriers trapped on the donors impurity, leading to nonequilibrium excess electron density in the conduction band [14]. So the intensity ratio between the exciton/trion on the $WS_2$ flake (Fig. 2(c)) also reflects the density distribution of the donor impurity. One can see that although the monolayer flake looks likely uniform in the optical microscope image and the total PL intensity image, the impurity density is different at different locations. Especially at the outside rings of the flake, which is more likely doped by the environment compared to the center.

*2.3 Exciton manipulation through varying the excitation density*

To further confirm the viewpoint of carrier photoionization and the trion generation, photoluminescence under different injected densities is carried out, as shown in Fig. 3. The PL spectrum shows single-Gaussian-peak below excitation density of $0.24kW/cm^2$ with resonance energy at 2.008eV (618.7nm), which is assigned to the radiative recombination bandage exciton [5]. While for higher intensities, a second emission peak appears on the low-energy side of the exciton emission and can be assigned to the emission of trion. The formation of trion requires excess electrons or holes, which mainly generate in two ways here: naturally n-doping inherited from the bulk crystal and photoionized carrier. The photoionization effect describes ionization of some neutral donor or acceptor impurities with energy level within the bandgap of monolayer $WS_2$. Free charges generating in photoionization can combine with an exciton to form a charge exciton (trion), as illustrated in Fig. 3(b). The photoionization effect is also observed in 4K monolayer WS$_2$ [14] and dominates the formation of trion.

Noticed that the trion peak shows a red-shift from 1.976eV to 1.972eV when the excitation density increases from 0.24kW/cm$^2$ to 204kW/cm$^2$, while the exciton peak is almost not change (or small red-shift below 1meV) in the whole measurement range. Thus the dissociation energy of trion △E, defined as the resonance energy difference between trion and exciton, increases with the rising of excitation power. The dissociation energy can also be wrote as the sum of the binding energy EX- and Fermi energy EF, namely, $\Delta E = E_{X^-} + E_F$ [14, 25]. The physical implication of this equation is that the free charge escaping from the dissociated trion needs to be released to above the Fermi surface where unoccupied states are available [21]. Thus the increasing △E is mainly due to the rising of Fermi energy and is also observed in other TMDs, including MoS$_2$ [13], WSe$_2$ [16], MoSe$_2$-WS$_2$ heterostructures [21] and low temperature monolayer (4K) WS$_2$ [14].

Fitting the curves in Fig. 3(a) with double-Gaussian-peak, this low-energy peak can be reliably separated from the main exciton peak. The fitting amplitudes are summarized in Fig. 3(c). The trion peak increases linearly with the excitation intensity in most of the measurement range, and only saturates at very high intensities (>60kW/cm$^2$). While the exciton peak is quickly saturated with the increasing excitation density and can be fitted well with the saturable absorption model [26].

$$I_{PL} \propto \frac{I \cdot I_s}{I + I_s} \quad (1)$$

where $I_{PL}$, $I$, $I_S$ are the PL intensity, excitation intensity and the saturation intensity, respectively. A saturation intensity of $I_S = 2.8$kW/cm$^2$ is obtained from the fitting process.

Fig. 3(d) summarizes the intensity ratio of the two carrier species (trion/exciton), which increases linearly with the rising excitation density with a slope of 0.043. In our measurement range from

0.24kW/cm$^2$ to 204kW/cm$^2$ at room temperature, this ratio can be continuously and linearly tuned from 0.1 to 7.7, indicating that the dominant carrier in the monolayer flake transfer from exciton to trion. The effective control of carrier species can be attributed to the high donor impurities density in the monolayer WS$_2$.

*2.4 Temperature dependent photoluminescence*

To gain more insight into of the generated exciton and trion in the monolayer WS$_2$, photoluminescence under different temperature is performed, as shown in Fig. 4. Here the excitation intensity is fixed to 0.1kW/cm$^2$ (0.7μW) to reduce the photoionization effect at room temperature. Cryogenic temperature is acquired by using the heating and cooling stage (LINKAM THMS600). As shown in Fig. 4(a), the PL spectrum above 260K show single-Gaussian line, which is assigned to neutral exciton. While below 260K, both the decreasing of thermal fluctuation and photoionization effect lead to the generation of trion. With decreasing of temperature, the trion peak increases significantly, especially below temperature of 180K. Fig. 4(b) shows the PL spectrum at 80K. Two flakes are measured under the same excitation power. Different PL spectrums but the same trend are acquired, indicating that the carrier species are the same, while the resonance energies of them are different, probably due to sample-to-sample variation. Fitting the PL spectrum of Flake 2, three peaks are acquired. Except for the exciton and trion, a third broad peak appears around 2.021eV and can be attributed to the defect related, trapped exciton states [7, 15]

Fig. 4(c) summarizes the fitting exciton and trion resonance energies as a function of temperature. With increasing temperature, both of them show a red-shift due to the red-shift bandgap induced by electron-phonon interaction. This red-shift can be described well by a standard semiconductor bandgap model [27]

$$E(T) = E(0) - S\hbar\omega(\coth(\hbar\omega/2k_BT) - 1) \qquad (2)$$

where *E(T)* is the bandgap at temperature of *T*, *S* is dimensionless coupling strength, $\hbar\omega$ is the phonon energy.

Fitting the resonance energy of exciton and trion with the above model. Some parameters are acquired: $S = 2.4$, $\hbar\omega = 31\text{meV}$ for both exciton and trion, and $E(0) = 2.063\text{eV}$ (2.040eV) for exciton (trion). Since the excitation intensity is very weak here, the binding energy nearly equals to dissociation energy. So here the binding energy of monolayer WS$_2$ is determined to be ~23meV at 0K. For other temperatures from 80 to 260K, the same fitting value of binding energy is acquired, as shown in the right column of Fig. 4(d), indicating that the binding energy of trion is independent to temperature.

The intensity ratio of trion and exciton can also be tuned effectively by varying temperature, as shown the left column of Fig. 4(d). When the temperature increases from 80K to 260K, the intensity ratio is tuned nonlinearly from 3.6 to 0.2 [18]. This effective tuning can be attributed to both the photoionization effect and the varying thermal fluctuation in monolayer WS$_2$.

## 3. Conclusion

In summary, the intrinsic excitonic properties of monolayer WS$_2$ have been studied. Both emissions from exciton and trion have been observed in ungated flakes at room temperature. Excitation intensity dependent PL shows the emissions of trion mainly due to the photoionization of donor impurity. Through varying excitation power, the intensity ratio of trion and exciton can be continuously and linearly tuned over a large range from 0.1 to 7.7. This intensity ratio can also be tuned from 0.2 to 3.6 by varying temperature from 260K to 80K, which is accompanied by red-shift of resonance energy of the exciton and trion due to electric-phonon coupling. The convenient and efficient strategy for manipulating the excitonic emission here is valuable to electronic, optoelectronic and valleytronics applications based on monolayer $WS_2$.

**Funding Inforamtion**

National Natural Science Foundation of China (NSFC) (61340017). The Scientific Researches Foundation of College of Optoelectronic Science and Engineering, National University of Defense Technology (No. 0100070014007).

**Acknowledgment**

We would like to thank Doctor Ye Tian (College of Mechatronics and Automation, National University of Defense Technology, Changsha, China) to provide the AFM image of monolayer $WS_2$.

**Figures**

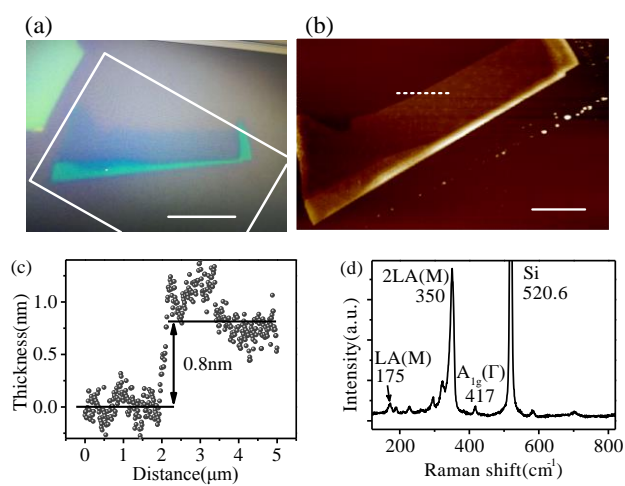

Fig. 1. Characterization of monolayer $WS_2$ (Flake 1). (a) Optical microscope image. The scale bar is 10μm. (b) AFM image of the area highlighted in (a). Scale bar, 5um. (c) AFM height profile along the dashed line in (b). (d) Raman spectrum of monolayer $WS_2$ at room temperature (295K).

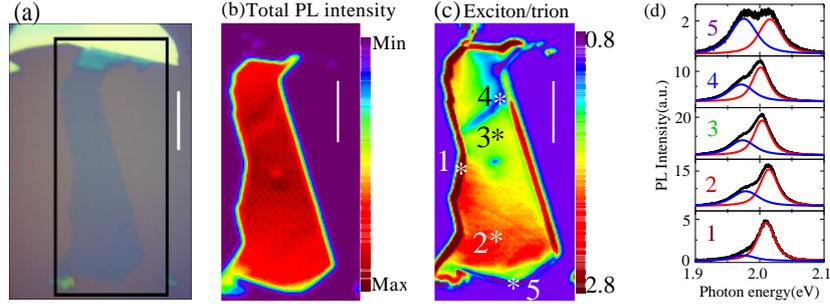

Fig. 2. Spatial maps of monolayer WS$_2$ (Flake 2). (a) Optical microscope image. (b) Total PL intensity (integrating from 600nm to 650nm) image and the intensity ratio of the PL from exciton (614~622nm) and trion (628~636nm). The white scale bar is 10μm for all figures. (c) The PL intensity ratio image of exciton (614~622nm) and trion (628~636nm). (d) PL spectrum at different locations marked in (c)

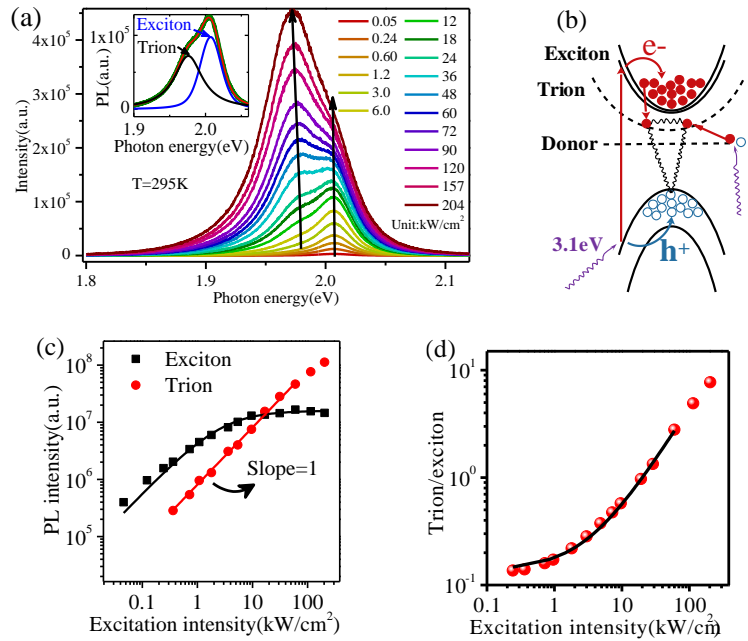

Fig. 3. (a) Excitation intensity dependent PL spectrum of monolayer WS$_2$ (Flake 1) at room temperature (295K). Two black arrows clarify the resonance energy evolution of exciton and trion. Inset shows a typical fitting result with double-Gaussian-peak at excitation intensity of 18kW/cm$^2$. (b) Schematics of the trion formation through photoionization of donor impurity. (c) PL intensity of the exciton (black square) and trion (red circle) as a function of excitation intensity. Black and red lines are the fits. (d) The intensity ratio of trion and exciton as a function of excitation intensity. Black line is the linear fit with equation expression of $I_{ratio} = 0.14 + 0.043I$.

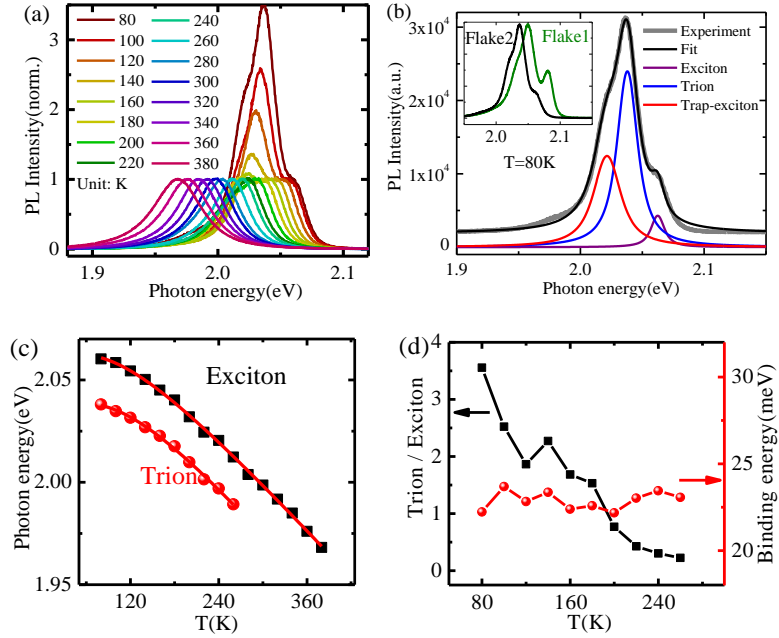

Fig. 4. (a) Normalized Photoluminescence spectra of WS$_2$ (Flake 2) under different temperatures, (b) Photoluminescence spectra of WS$_2$ at 80K, the color lines are three-Gaussian-peak fits which can be assigned to exciton (purple), trion (blue) and trap state exciton (red), respectively. Inset shows photoluminescence spectra of two flakes used in this paper at 80K. (c) The resonance energy of exciton (black square) and trion (red sphere) as a function of temperature. Lines are fitting curves with a standard semiconductor bandgap model. (d) The intensity proportion of trion/exciton (black square) and binding energy of trion (red sphere) at different temperatures.